# $Fe_3O_4$(001) films on Fe(001) - termination and reconstruction of iron rich surfaces

N. Spiridis,[1] J. Barbasz,[1] Z. Łodziana,[2,3] and J. Korecki.[1,4]

[1]Institute of Catalysis & Surface Chemistry, Polish Academy of Sciences, 30-239 Kraków, Poland,

[2]Center for Atomic-scale Materials Physics, DTU, 2800 Lyngby, Denmark,

[3] Henryk Niewodniczański Institute of Nuclear Physics, Polish Academy of Sciences, 31-342 Kraków, Poland,

[4]Faculty of Physics & Applied Computer Science, AGH University of Science and Technology, 30-059 Kraków, Poland.

High-quality and impurity-free magnetite surfaces with $(\sqrt{2}\times\sqrt{2})R45^o$ reconstruction have been obtained for the $Fe_3O_4$(001) epitaxial films deposited on Fe(001). Based on atomically resolved STM images for both negative and positive sample polarity and Density Functional Theory calculations, a model of the magnetite (001) surface terminated with Fe ions forming dimers on the reconstructed $(\sqrt{2}\times\sqrt{2})R45^o$ octahedral iron layer is proposed.

68.47.Gh, 68.37.Ef, 68.55.-a

## I. INTRODUCTION

Controversies about structural details of the $Fe_3O_4(001)$ polar surface,[1] constituting one of the possible low-index magnetite terminations, remain unsolved since the very first scanning tunneling microscopy (STM) experiment on this surface.[2] Magnetite ($Fe_3O_4$) crystallizes in the inverse spinel structure with a lattice constant of 8.40 Å. Fe ions are located at two different interstitial sites octahedrally and tetrahedrally coordinated to oxygen that forms a close-packed cubic structure. The tetrahedral sites (A) are occupied by trivalent Fe ions, whereas a randomly arranged mixture of the tri- and divalent Fe ions occupies the octahedral sites (B). At T~125 K, magnetite undergoes a metal-insulator transition, known as the Verwey transition and commonly interpreted as a long-range electron charge ordering in the octahedral Fe sublattice. For a review of magnetite properties see Refs. 3, 4.

The structure of the (001) magnetite surface has been intensively studied for single crystals[2, 5, 6, 7, 8] as well as for epitaxial films.[9, 10, 11, 12, 13, 14, 15] The surface is usually discussed as being composed of atomic sublayers, containing either only tetrahedral iron ions Fe(A) (the so-called A-layer) or oxygen and octahedral iron ions Fe(B) (the so-called B-layer). The distance between A or B layers is about 2.1 Å, whereas the smallest interlayer (A-B) spacing is about 1.1 Å. The most typical reconstruction seen by then low energy electron diffraction (LEED) for the (001) surface is ($\sqrt{2}\times\sqrt{2}$)R45° (corresponding to the 8.40x8.40 Å$^2$ surface mesh) as related to the (1x1) primitive surface unit cell with the 5.94x5.94 Å$^2$ periodicity of the bulk terminated surface. Neither A nor B bulk termination of the $Fe_3O_4(001)$ surface is charge-compensated, and a number of models assume that the charge neutrality condition is a driving force behind the reconstruction. The obvious way to achieve the autocompensated $Fe_3O_4(001)$ surface with the observed reconstruction is to remove certain surface atoms: half of $Fe^{3+}$ ions for the A termination[10, 11, 12, 14] or a number of oxygens for the B termination[10,13]. However, surface stability can be achieved also through electronic degrees of freedom.[1] Thus, models with full A- or B-type layer termination with a specific surface electronic and geometric structure were also proposed.[2,5,7,8,9] The interpretation was obscured by ambiguity of the surface stoichiometry since there were no procedures to control whether the surface layer was oxygen- or iron-rich. Both single crystal and epitaxial film surfaces for Ultra High Vacuum (UHV) studies are prepared in processes occurring in a broad range of temperature and oxygen partial pressures. Consequently, there is no consistency between the variety of the proposed models and the real space images of the $Fe_3O_4$(001) surfaces obtained by STM.[2,5,8,9,13,15]

The reported STM images of the $Fe_3O_4$(001) surface differ in many details, but one feature is common for all observations. On flat terraces, which are terminated with steps of 2.1 Å in height, atomic rows spaced by 6 Å are seen at positive sample biases. These are mutually perpendicular on neighboring terraces. The step height corresponds to the A-A or B-B layer spacing. Due to symmetry and spacing of the rows, they are commonly attributed to Fe ions in octahedral sites. The occupied state images are rarely reported. They reveal elongated shapes forming a square 8.4Åx8.4 Å mesh, interpreted as clusters of atoms within the tetrahedral termination.[5, 9] To our knowledge, images of the empty and filled states have never been observed for the same sample with the ($\sqrt{2} \times \sqrt{2}$)R45° reconstruction.

In the present study, we propose an alternative method of a preparing $Fe_3O_4$(001) surface with a stable and reproducible ($\sqrt{2} \times \sqrt{2}$)R45° reconstruction by depositing magnetite on a Fe(001) film. The surface gives STM images of high stability and atomic resolution for negative and positive sample polarity, allowing us to image details of the surface structure and, when combined with Density Functional Theory (DFT) calculations, to propose a model of the iron rich surface. We show that using different preparation methods the magnetite surface with octahedral (B) or partially filled and tetrahedral termination with Fe dimers can be stabilized.

II. METHODS AND SAMPLE PREPARATION

The magnetite films were obtained in a UHV system on cleaved MgO(001) substrates. Two preparation schemes were used. The surface structure of $Fe_3O_4$(001) films prepared in a „classical" way, directly on MgO, by oxygen assisted deposition of Fe at 250 $^o$C, was described in details in our earlier papers.[15,16] The alternative preparation method consisted of depositing a 200 Å buffer layer of epitaxial Fe(001) on the MgO(001) substrate to guarantee an iron reservoir to balance the magnetite stoichiometry. On the Fe(001) surface an oxide layer was formed by oxygen-assisted deposition of Fe at $10^{-4}$ Pa $O_2$ on the substrate, which was held at 250 $^o$C. Finally, the samples were annealed at 500 $^o$C for 60 minutes. The as-prepared films were characterized *in situ* by Auger electron spectroscopy (AES) and LEED, *ex situ* by the conversion electron Mössbauer spectroscopy, and display typical $Fe_3O_4$(001) features. The ($\sqrt{2}$x$\sqrt{2}$)R45$^o$ reconstruction remained stable also after the post-preparation annealing, whereas the films deposited directly on MgO(001) showed, after the same treatment, major changes of the surface structure induced by the diffusion of Mg from the substrate into the magnetite film.[13,15] Apparently, the Fe(001) layer sets an effective barrier for Mg diffusion.

The STM measurements were carried out *in situ* with a room-temperature STM head (Burleigh) using electrochemically etched tungsten tips. The corresponding theoretical simulations were based on DFT calculations. The plane wave pseudopotential method with the gradient corrected exchange-correlation functional[17] was used.[18] The Monkhorst-Pack k-point sampling mesh with density of $0.1 A^{-1}$, together with a kinetic energy cutoff of 340 eV and the charge density grid of 680 eV were applied. The electronic density was determined by iterative diagonalization of the Kohn-Sham Hamiltonian, and the resulting Kohn-Sham eigenstates were populated according to the Fermi statistics with a finite temperature smearing of kT=0.015 eV. The calculations reproduce the magnetic and structural properties of the bulk magnetite[3] very well, giving the lattice constant $a_0$=8.39 Å and magnetic moments of 3.34 $\mu_B$ for Fe(B) and 3.28 $\mu_B$ for Fe(A). The surface calculations were performed in a stoichiometric slab geometry with eight atomic layers in the reconstructed $(\sqrt{2}\times\sqrt{2})R45^o$ unit cell and with six atomic layers in p(2x2) geometry to asses stability of Fe dimers on the surface. The vacuum region extended up to 16 Å and the dipole correction was applied. The STM topographs in the constant current (electron density) mode were simulated based on the Tresoff-Hamann formalism.[19]

III. RESULTS AND DISCUSSION

Large-scale in situ STM images for $Fe_3O_4$(001) surfaces prepared on Fe(001) epitaxial films (an example is shown in Fig. 1a) reveal surface topography similar to that observed for well-ordered single crystalline $Fe_3O_4$(001) surfaces.[8] The high-temperature post-deposition annealing results in flat terraces, extending over 30x30 $nm^2$ in average. The terraces are separated by 2 Å high atomic steps along the <110> directions, terminated by a pair of screw dislocations. For comparison, the STM image of a $Fe_3O_4$(001) film deposited in the classical way, directly on MgO, is presented in Fig. 1b. The limitation of deposition/annealing temperature, as discussed above, is reflected in the surface topography – the much smaller terraces and the large number of defects. Atomically resolved images could be obtained in a broad range of positive and negative bias voltages for the magnetite films deposited on Fe(001)/MgO(001). The most typical ones, appearing reproducibly for all investigated samples (over 20) are shown in Fig. 2. At the positive sample bias, a network of regularly arranged dark features with the 8.4x8.4 Å$^2$ periodicity corresponding to the ($\sqrt{2}$x$\sqrt{2}$)R45$^o$ reconstruction is seen between the bright rows along the <110> directions spaced by 6 Å (Fig.2a,c). The rows are mutually perpendicular on neighboring terraces. The sample bias can be reproducibly changed to negative values (typically $V_s$ = -1.7 V), yielding entirely different images dominated by well-separated bright protrusions with locally different surface densities (Fig.2b). The protrusions, which we call ovals, are elongated in the <110> directions and show some internal structure as seen at higher magnification. The ovals form rows along the <110> directions that are spaced by 6 Å, while the distance between the ovals in a row is 12 Å. Consequently, in the densest packed areas, ovals form a 8.4x8.4 Å$^2$ mesh (the circle II in Fig. 2b). Every other row of ovals is frequently missing, and thus the areas with a 12x12 Å$^2$ mesh are formed (the circle III in Fig. 2b). The oval statistics over 2500 nm$^2$ taken for many terraces and samples gives the average occupation of 0.50±0.03 oval per the unit cell of the ($\sqrt{2}$x$\sqrt{2}$)R45$^o$ reconstruction The ovals, apart from their densities, are very similar to the shapes observed by Tarrach *et al.*[5] and Gaines *et al.*[9] for single-crystalline and thin-film surfaces, respectively. Gaines et al.[9] described them as diffused clusters of atoms, which may consist of both tetrahedral and octahedral Fe ions, and are difficult to interpret due to a complicated atomic arrangement.

The exceptional stability of the STM images allowed us to change sample polarity during the scan (indicated by the arrow in Fig. 2c), and thus the respective positions of atomic scale details in the pair of images could be easily identified as shown by oval contours in Fig. 2c and Fig. 2d. The determination of the absolute positions of the atomic details with respect to the (1x1) surface unit cells done with an assumption that any row-like features appearing along <110> are due to octahedral irons, as for a bulk-like surface. Within such a hypothesis the position of the ovals is localized between the octahedral irons, which suggests that they appear predominantly due to tunneling from tetrahedral Fe ions. Consequently, we infer that the surface is terminated with a partially occupied layer of the tetrahedral Fe(A) ions. The occupied state atomic images presented in Figs 2b,d are not the only ones characteristic for the surfaces of $Fe_3O_4$(001) on Fe(001). At a special tip state,[20] enhanced electronic contrast and resolution could be achieved with some experimental effort. A corresponding example is shown in Figs. 3a,b. The ovals, showing only a minor splitting at $V_s$ = -1.85 V in Fig. 3b, become well resolved when the bias voltage is changed to $V_s$= -0.73 V (Fig. 3a), suggesting that they originate from two different atoms, henceforth referred to as a dimer. As seen from the height profiles in Fig. 3, the distance between the dimer centers, as well as between the ovals, is 12 Å, whereas the atoms in a dimer are spaced by 4.9 Å, as compared to the 5.96 Å distance between the tetrahedral $Fe^{3+}$ ions in the ideal A-layer. Moreover, the structure of rows along <110> spaced by 6 Å becomes very distinct.

As shown schematically in Fig. 2c, a single oval covers two tetrahedral Fe positions and hence the statistics of the ovals (or dimers) observed on the $Fe_3O_4$(001) surfaces prepared on Fe(001) films yields a half-monolayer of Fe(A). Therefore, we conclude that on average the surface is terminated with ½ monolayer of Fe(A) ions, which form dimers along <110>. The 50% occupation is realized on a large scale, whereas small areas without dimers (i.e. B terminated), with (2x2)-12x12 Å$^2$ and with ($\sqrt{2}$x$\sqrt{2}$)R45°-8.4x8.4 Å$^2$ square dimer mesh can coexist locally, as marked in Fig. 2b.[21] For surface neutrality, it is enough that the charge is compensated on the large scale by a combination of small areas with three different terminations: full A-layer, half A-layer and B-layer (presumably near defects). However, such a terminating layer, with a short-range order only, cannot be responsible for the almost perfect ($\sqrt{2}$x$\sqrt{2}$)R45° reconstruction observed in the empty state images (Fig.2 a,c). We believe that the reconstruction comes from the outermost surface B-layer, and the arrangement of the Fe(A) dimers in the terminating layer only reflects the reconstruction.

The two preparation methods of $Fe_3O_4$(001) films, the "classical" one, directly on MgO(001) and the present one, on a Fe(001) buffer layer, result in surfaces that show the same ($\sqrt{2}$x$\sqrt{2}$)R45° reconstruction in LEED but differ in AES and STM. The classical preparation gives us the Auger signal ratio of the 510 eV oxygen and 651 eV iron lines $R = 3.34(5)$, very close to that reported by Ruby et al.[22] For the magnetite films on Fe, the $R$ value is reduced down to 2.96(5), indicating an iron rich termination, which we also postulate based on the STM analysis. Contrary to the magnetite samples grown on the Fe buffer layer, we have never observed the oval features for the classical preparation. In this case, with some experimental effort, we were also able to acquire STM images from the same areas at different polarities (Fig. 4). For both polarities the images are dominated by the structure of rows separated by about 6 Å. As we discussed earlier,[15] at certain bias voltages the rows have typical modulation that is responsible for the ($\sqrt{2}$x$\sqrt{2}$)R45° reconstruction. Clearly, our observations support the earlier findings that such a surface is terminated with the oxygen-rich layer and the models of reconstructed B termination apply.[8,13]

A number of recent papers have tried to explain theoretically the structure of the (001) magnetite surface.[23, 24, 25] In particular, Pentcheva et al.[25] explained the ($\sqrt{2}$x$\sqrt{2}$)R45° reconstruction of the B-terminated surface in terms of the Jahn-Teller distortion. Our present calculations of the B termination agree well with this picture. The modification of the local spin moments on the surface Fe(B) gives rise to the ordering of octahedral cations: pairs of cations along the [110] direction with $\mu_{Fe(B1)}$=3.50 $\mu_B$ and $\mu_{Fe(B2)}$=2.74 $\mu_B$, are formed, similar to that discussed by Shvets et al.[23] for the (001) surface of a bulk magnetite crystal. The changes of the magnetic moments on Fe(B) cations are related to the ordering of electrons in the $d_{x^2-y^2}$ atomic orbitals. The calculated STM images render the surface reconstruction but oval-like shapes between octahedral rows were not identified at any bias voltage. It supports our experimental conclusion that the termination of our magnetite films on the Fe (001) buffer film must be something else than the B-layer.

Due to limited computer resources we considered simplified models of A-terminated surface: the termination with Fe(A) dimers on ($\sqrt{2}$x$\sqrt{2}$)R45° that corresponds to coverage of 1 ML of Fe(A). The separate calculations were performed for iron dimers on a p(2x2) surface, which reflects 0.5ML coverage of Fe(A) on the surface unit cell. Surfaces with Fe(A) arranged in their regular lattice sites were also considered for calculations of the surface stability. At the equilibrium, Fe-dimers are located ~0.35 Å above the surface plane (defined by oxygen) for all considered geometries. In the optimized geometry, the Fe(A) cations in a dimer are separated by distance not shorter than d=2.90 Å. This is significantly longer than d < 2Å for a Fe$_2$ dimer in vacuum. The geometry of the Fe(A) dimers on the p(2x2) surface varies depending on dimer location. For 0.5 ML coverage, both elongated dimers with d=3.74 Å and with d=2.90 Å are stable. The ($\sqrt{2}$x$\sqrt{2}$)R45° surface terminated by the Fe-dimers is more stable than the surface with Fe(A) regularly arranged on top of the B-layer. Detailed comparison of the dimer stability on the p(2x2) surface, calculated within generalized surface energy formalism,[26] indicated that the stability of the Fe-dimers is comparable (within 3meV/Å$^2$) to the stability of reconstructed surface termination with half A-layer, reported previously.[24,25] For the reconstructed 0.5ML Fe(A) termination, cations are incorporated into the surface layer, similarly as for the more stable termination reported in Ref. [25], while the Fe-dimers always stay above the surface. Calculated stability of the p(2x2) surface with the Fe-dimers indicates that there are no thermodynamic restrictions against the existence of this surface under UHV conditions. The theoretical STM picture of the dimeric A-surface (Fig. 3c) reproduces well the experimental observation for negative sample bias as shown by the theoretical profile along <110> (the gray line in the plots of Fig. 3). However, the full diversity of STM images for all experimental bias voltages could not be well reproduced. In particular, the oval shapes dominate also in theoretical STM images of the empty states. The reason could be a strong on-site Coulomb interaction that further splits the occupied and empty Fe(A) states.[27] It is also possible that the images have been obtained with an impurity atom (eg. oxygen) attached to the tip. These effects are not included in our calculations. The problem requires further investigations, possibly taking into account alternative models of the Fe rich termination.

According to recent DFT calculations,[24,25] the Fe-rich termination discussed above is not the most favorable energetically even under oxygen-poor conditions, which remains in striking opposition to our experimental results. We note, however, that the bulk phase diagram of the $FeO$ - $Fe_3O_4$ - $Fe_2O_3$ system[3] indicates stability of magnetite for much lower oxygen partial pressure than that taken in Ref. 25 as "O-poor limit". Moreover, the short-range order of Fe(A) and dimerization on the surface is not considered in the models of Refs. 24,25.

## IV. CONCLUSIONS

During the decade many controversies have arisen concerning the $Fe_3O_4$(001) termination and reconstruction. The surface was shown to be very sensitive to preparation method. Apparently conflicting STM data probably concern surfaces that are differently influenced by impurity segregation or by deviations from stoichiometry due to the reducing or oxidizing procedures applied when preparing or recovering the surface. We have proposed the alternative method of preparation, in which $Fe_3O_4(001)$ films are grown on Fe(001)/Mg(001) substrates. We have shown that the surface structure of such films has a different termination than that of films grown on MgO(001) in a classical way, despite the similar reconstruction seen in low- energy electron diffraction. The analysis of the high-resolution images acquired for both negative and positive sample vs. tip polarities combined with DFT calculations leads to the model of the Fe rich surface, which explains and unifies many features of the surface previously postulated. In particular, complexity and atomic details of the Fe(A) terminated surface, which can be obtained under Fe-rich preparation conditions, have been addressed. Moreover, we have shown how the different preparation conditions – oxygen rich vs. iron rich, alter the oxide surface on the atomic scale. This opens possibilities for comparative studies of surface adsorption and reactivity.

ACKNOWLEDGEMENTS

This work was supported by the European Community under the Specific Targeted Research Project Contract No. NMP4-CT-2003-001516 (DYNASYNC) and by the Polish Ministry of Education and Science. J.K. acknowledges the Foundation for Polish Science (FNP) for support. Z.Ł. acknowledges ICM grant No.G28-22. The authors thank Dr. N. Bailey for reading the manuscript.

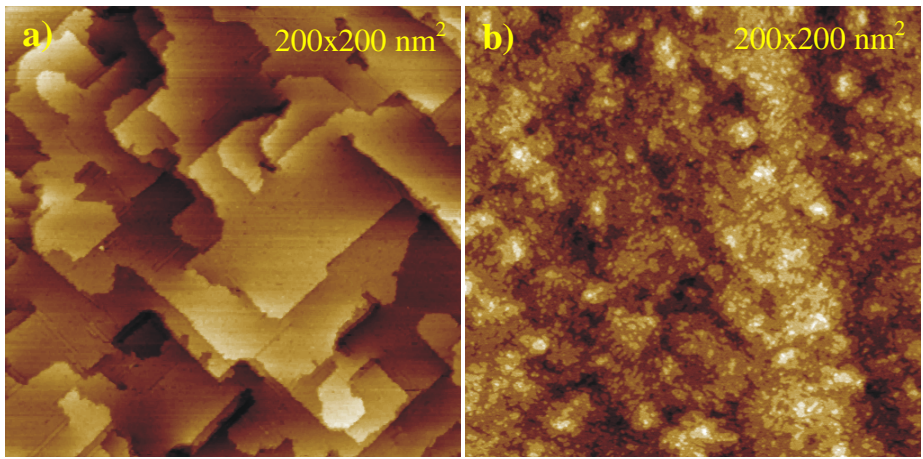

**Fig. 1 (color on line)** Topographic (constant current) STM images for the $Fe_3O_4(001)$ surface of the epitaxial magnetite film on Fe(001)/MgO(001) (a) and on MgO(001) (b). The images are taken at the positive sample bias of about 0.8 V. For the details of preparation conditions see text.

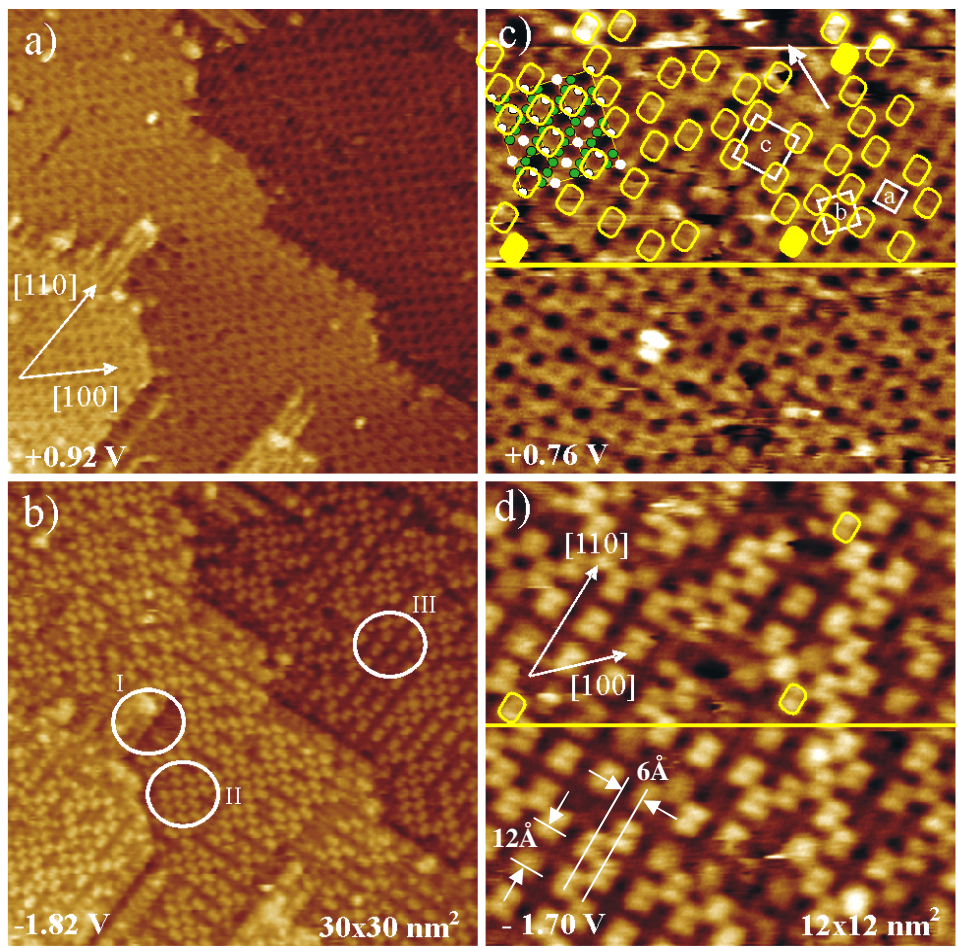

**Fig. 2 (color)** Constant current STM images of the empty (top) and occupied (bottom) states for the Fe$_3$O$_4$(001) surface of the epitaxial magnetite film on Fe(001)/MgO(001). Scans from the same areas are shown in columns. The circles in b) indicate the areas without dimers (I), with ($\sqrt{2}\times\sqrt{2}$)R45° - 8.4x8.4 Å$^2$ (II), and with (2x2) -12x12 Å$^2$ (III) square dimer mesh, respectively. The arrow in c) indicates the line where the sample bias was reversed. The contours of ovals seen in d) are superimposed on the upper part of c), and the equivalent positions are marked by corresponding filled and empty contours. Squares labeled with *a*, *b* and *c* represent unit cells of (1x1), ($\sqrt{2}\times\sqrt{2}$)R45° and (2x2) reconstructions, respectively.

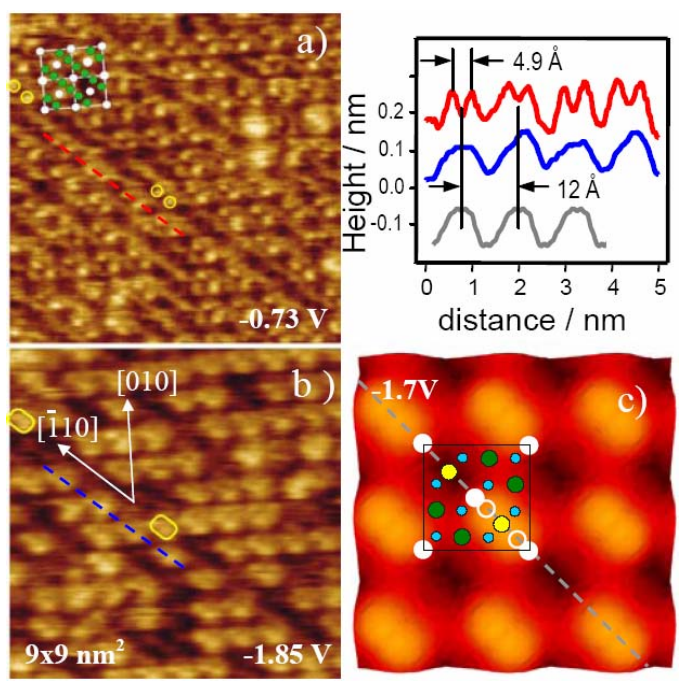

**Fig. 3 (color)** STM constant current scans taken subsequently from the same area at -0.73V (a) and -1.85V (b) sample bias and theoretical STM image of the dimeric A termination at negative ($V_s$=-1.7 V) sample bias (c). Small circles in a) mark positions of atoms that are equivalent to oval contours in b). The experimental height profiles along <110>, with colors corresponding to lines marked in the a) and b), and the theoretical one in gray are shown in upper right plot. Circles models in a) and c) indicate atomic positions on an ideal surface with tetrahedral termination. Green circles represent surface octahedral Fe atoms arranged in rows along <110>, white circles are surface tetrahedral Fe atoms in their bulk-like positions arranged in the (1x1) square lattice, open white circles are surface tetrahedral irons in dimers, yellow circles are sub-surface tetrahedral Fe atoms and small blue circles are oxygen atoms.

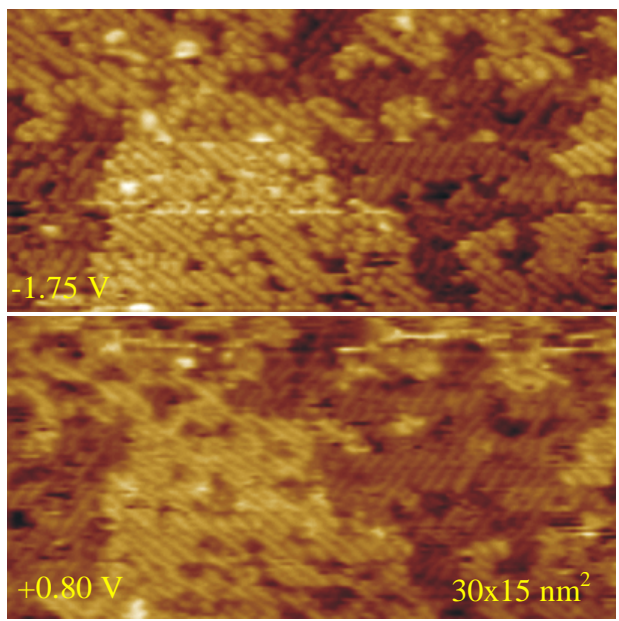

**Fig. 4 (color on line)** Constant current STM images of the same surface area for the Fe$_3$O$_4$(001) epitaxial magnetite film on MgO(001) taken at a negative (top) and positive (bottom) sample bias.

[1] C. Noguera, J. Phys.: Condens. Matter **12**, R367 (2000).

[2] R. R. Wiesendanger, I. V. Shvets, D. Bürgler, G. Tarrach, H. J. Güntherodt, J.M.D. Coey, S. Gräser, Science **255,** 583 (1992).

[3] F. Walz, J. Phys.: Condens. Matter **14**, R285 (2002).

[4] J. Garcia, G. Subias, J. Phys.: Condens. Matter **16**, R145 (2004).

[5] G. Tarrach, D. Burgler, T. Schaub, R. Wiesendanger, H.-J. Güntherodt, Surf. Sci. **285**, 1 (1993).

[6] J.M.D. Coey, I. V. Shvets, R Wiesendanger, H-J. Güntherodt, J. Appl. Phys. **73**, 6742 (1993).

[7] R. Koltun, M. Herrmann, G. Güntherodt, V.A.M. Brabers, Appl. Phys. A **73**, 49 (2001).

[8] G. Mariotto, S. Murphy, I. V. Shvets, Phys. Rev. B **66**, 245426 (2002).

[9] J.M. Gaines, P.J.H. Bloemen, J.T. Kohlhepp, C.W.T. Bulle-Lieuwma, R.M. Wolf, A. Reinders, R.M. Jungblut, P.A.A. van der Heijden, J.T.W.M. van Eemeren, J. aan de Stegge, W.J.M. de Jonge, Surf. Sci. **373**, 85 (1997); J.M. Gaines, J.T. Kohlhepp, J.T.W.M.v. Eemeren, R.J.G. Elfrink, F. Roozeboom, W.J.M.d. Jonge, Mater. Res. Soc. Symp. Proc. Vol. **474**, 191 (1997).

[10] F.C. Voogt, Ph.D. Thesis, Departments of Chemical Physics and Nuclear Solid State Physics, University of Groningen, Netherlands, 1998. F.C. Voogt et al., Phys. Rev. B **60**, 11193 (1999).

[11] S.A. Chambers, S.A. Joyce, Surf. Sci. **420**, 111 (1999).

[12] S.A. Chambers, S. Thevuthasan, S.A. Joyce, Surf. Sci. **450**, L273 (2000).

[13] B. Stanka, W. Hebenstreit, U. Diebold, S.A. Chambers, Surf. Sci. 448, 49 (2000).

[14] A.V. Mijiritskii, D.O. Boerma, Surf. Sci. **486**, 73 (2001).

[15] N. Spiridis, B. Handke, T. Slezak, J. Barbasz, M. Zajac, J. Haber, J. Korecki, J. Phys. Chem. B **108**, 14356 (2004).

[16] J. Korecki, B. Handke, N. Spiridis, T. Ślęzak, I. Flis-Kabulska, J. Haber, Thin Solid Films **412**, 14 (2002).

[17] J.P. Perdew, J.A. Chevary, S.H. Vosko, K.A. Jackson, M.R. Pederson, D.J. Singh, C. Fiolhais, Phys. Rev. B **46** 6671 (1992).

[18] http://www.fysik.dtu.dk/campos/Dacapo/.

[19] J. Tersoff and D.R. Hamann, Phys. Rev. Lett. **50** 1998 (1983).

[20] M. Schmid, H. Stadler, and P. Varga, Phys. Rev. Lett. **70**, 1441 (1993).

[21] The small areas with (2x2) reconstruction give rise to weak and broad LEED spots visible at the electron energy as low as 20 eV.

[22] C. Ruby, J. Fusy and J.-M.R. Génin, Thin Solid Films **352**, 22 (1999).

[23] I.V. Shvets, G. Mariotto, K. Jordan, N. Berdunov, R. Kantor, S. Murphy, Phys. Rev. B **70**, 155406 (2004).

[24] C. Cheng, Phys. Rev. B **71**, 052401 (2005).

[25] R. Pentcheva, F. Wendler, H.L. Meyerheim, W. Moritz, N. Jedrecy, M. Scheffler, Phys. Rev. Lett. **94**, 126101 (2005).

[26] K. Reuter, M. Scheffler, Phys. Rev.B **65**, 035406 (2002)

[27] H.-T. Jeng, G. Y. Guo, and D. J. Huang, Phys. Rev. Lett. **93**, 156403 (2004).